# Double-Side Cocatalytic Activation of Anodic $TiO_2$ Nanotube Membranes with Sputter-Coated Pt for Photocatalytic $H_2$ Generation from Water-Ethanol Mixtures

Gihoon Cha,[a] Marco Altomare,*[a] Nhat Truong Nguyen,[a] Nicola Taccardi,[b] Kiyoung Lee,[c] and Patrik Schmuki*[a][d]

[a] G. Cha, Dr. M. Altomare, N.T. Nguyen, Prof. Dr. P. Schmuki
Department of Materials Science and Engineering WW4-LKO
University of Erlangen-Nuremberg
Martensstrasse 7, Erlangen, D-91058, Germany
*Corresponding Authors.
Email: schmuki@ww.uni-erlangen.de; marco.altomare@fau.de

[b] Dr. N. Taccardi
Lehrstuhl für Chemische Reaktionstechnik
University of Erlangen-Nuremberg
Egerlandstraße 3, 91058 Erlangen, Germany

[c] Prof. Dr. K. Lee
Department of Energy Chemical Engineering, School of Nano & Materials Science and Engineering
Kyungpook National University, South Korea

[d] Prof. Dr. P. Schmuki
Department of Chemistry
King Abdulaziz University
Jeddah, Saudi Arabia

**Abstract:** Self-standing $TiO_2$ nanotube layers in the form of membranes are fabricated by self-organizing anodization of Ti metal and a potential shock technique. The membranes were then decorated by sputtering different Pt amounts *i)* only at the top, *ii)* only at the bottom or *iii)* at both top and bottom of the tube layers. The Pt-decorated membranes are transferred either in tube top up or in tube top down configuration onto FTO slides and investigated after crystallization as photocatalysts for $H_2$ generation using either front or back-side light irradiation. Double-side Pt-decoration of the tube membranes leads to higher $H_2$ generation rates (independent of tube and light irradiation configuration) compared to membranes decorated at only one side with similar overall Pt amounts. The results suggest that this effect is not ascribed to the overall amount of Pt cocatalyst as such but rather to its distribution at both tube extremities. This leads to optimized light absorption and electron diffusion/transfer dynamics: the central part of the membranes act as light harvesting zone and electrons therein generated can diffuse towards the Pt/$TiO_2$ active zones (tube extremities) where they can react with the environment and generate $H_2$ gas.

## Introduction

Hydrogen is nowadays regarded as promising energy vector of the future and many efforts have been devoted to identify reliable and scalable catalytic processes that allow for its green generation from cheap and abundant sources.[1–4]

A most ideal way to produce $H_2$ is by photocatalytic splitting of water using an adequate semiconductor under sunlight irradiation.[5] Briefly, a photocatalytic process is based on light absorption by a semiconductor ($h\nu \geq E_g$) and subsequent generation of charge carriers (electrons and holes in the conduction and valence bands, respectively), and their reaction with the environment. That is, the charge carriers can reach the semiconductor surface and react with redox species in the environment, if the band energy position and red-ox states are suitable (in terms of electrochemical potentials).[6–8]

Since the work of Fujishima in 1972,[9] $TiO_2$ has become the most investigated metal oxide semiconductors for photocatalytic applications owing to its stability, low cost, large availability,[10–13] and most importantly for its band energy positions that are suitable to split water into $H_2$ and $O_2$.[14–16] However, $TiO_2$ is commonly used to generate $H_2$ from water-organic mixtures, where the organic, *e.g.*, methanol, ethanol, *etc.*, act as efficient hole scavenger (undergoing oxidation into $CO_2$) – this typically reduces charge carrier recombination and leads to higher $H_2$ generation rates.[17,18]

Besides the use of a hole scavenger, another common strategy to improve the efficiency of a photocatalytic process is to use nanostructured $TiO_2$-based materials such as nanoparticles, nanowires and nanorods. Owing to their large specific surface area, and thus to the facilitated charge transport towards the semiconductor surface (*i.e.*, short $h^+$ and $e^-$ diffusion length) nanostructured materials typically lead to improved photocatalytic efficiencies.[19–24]

Among several methods to synthesize $TiO_2$ nanostructures, the anodization of Ti metal under self-organizing electrochemical conditions to form highly-ordered arrays of $TiO_2$ nanotubes (NT) has gained large attention in the last decades.[25] A most relevant advantage of this process is that simple electrochemical conditions (temperature, anodization voltage, electrolyte composition) can be adjusted to form NT layers with precise physicochemical and morphological features in order to maximize their photocatalytic activity.[26–30]

Nevertheless, the photocatalytic $H_2$ evolution reaction on $TiO_2$ is hindered by a sluggish charge transfer kinetic. Thus, to reach reasonable $H_2$ evolution rates, the surface of the photocatalyst needs to be modified by deposition of noble metal cocatalytic nanoparticles (NP) (Pt, Au, Pd).[17,31–36] The noble metal at the $TiO_2$ surface forms a Schottky-type junction that typically improves the photocatalytic efficiency by trapping conduction band electrons and by mediating their transfer to the reactants in the environment, *e.g.*, $H_2O$. In other words, the noble metal limits the charge recombination *phenomena.*[6,37–43] Additionally, some noble metals, namely Pt, also aid the recombination reaction of $H^0$ to form $H_2$ gas.[44,45]

Size, location and decoration density of the cocatalyst NPs are key factors towards a photocatalytic enhancement. Particularly for one-dimensional (1D) nanostructures such as $TiO_2$ NTs it has been shown that the cocatalyst/$TiO_2$ interface can be designed in order to satisfy a set of critical factors such as the particle density and formed local Schottky junctions (which affect the electronic structure of $TiO_2$), the free $TiO_2$ surface area *vs.* that shaded by metal particles, the nature of metal/oxide junction and oxide/environment interface.

An example of photocatalytically efficient Pt/$TiO_2$ geometry was reported by Nguyen *et al.*[46] In this work cocatalytic Pt NPs were sputter-deposited only at the upper part (mouth) of anodic $TiO_2$ NT arrays. This site-selective decoration more likely induces a gradient of the semiconductor Fermi level ($E_F$) in the tube walls along the length of each $TiO_2$ tube. Thus, the transfer of electrons that are generated deep in the $TiO_2$ tube layers towards the metal/$TiO_2$ coupled zone (photocatalytically active zone) is facilitated by the formed junction, that is, by the gradient of $E_F$ along the $TiO_2$ tube walls. In other words, such geometry provides an electron harvesting zone (light penetration depth can be of a few 100 nm up to a few µm deep in the tube structure[47]) combined with a charge-transfer zone (tube top).[48]

In this work we fabricate highly-ordered $TiO_2$ nanotube membranes by electrochemical anodization (for the tube growth) and a potential shock technique (to detach the NT layers from the Ti substrate and form self-standing tube layers).[49,50] We then decorated these tube membranes with Pt (cocatalyst) by sputter-coating only at their top, only at their bottom, or at both sides. After the cocatalyst deposition, the Pt/$TiO_2$ NT membranes are transferred onto FTO glasses (either in tube top up or tube top down configuration).

The Pt/$TiO_2$ tube membranes on FTO glasses are then crystallized and investigated as photocatalysts for $H_2$ generation from water-methanol mixtures under UV light irradiation ($\lambda$ = 325 nm). We particularly explore the effects of the Pt amount and location (only top, only bottom, both top and bottom), configuration of the tube layers (tube top up or down) and light irradiation pathway (*i.e.*, front- or back-side irradiation) on the photocatalytic $H_2$ generation efficiency.

We find that the deposition of Pt at both the extremities (tube top and bottom) leads to higher $H_2$ generation rates compared to membranes decorate with similar Pt amounts at only one side (either top- or bottom-side decoration). The photocatalytic enhancement should not be ascribed to only the overall Pt cocatalyst amount as such, but also to the double-side decoration that creates an inner zone of the NTs acting as light absorber and providing charge carriers (electrons) to the Pt/$TiO_2$ coupled zones at the tube extremities (photocatalytically active zones) where the $H_2$ generation takes place. Thus, on the one hand the $TiO_2$ NT membranes grant a full light absorption and on the other hand the double-side Pt decoration leads to efficient electron collection (trapping) and transfer to the environment for a maximized $H_2$ evolution.

## Results and Discussion

Scheme 1 shows the process (Scheme 1a) used in this work to fabricate Pt/$TiO_2$ NT membranes on FTO glass and their arrangement for studying the photocatalytic activity for $H_2$ generation.

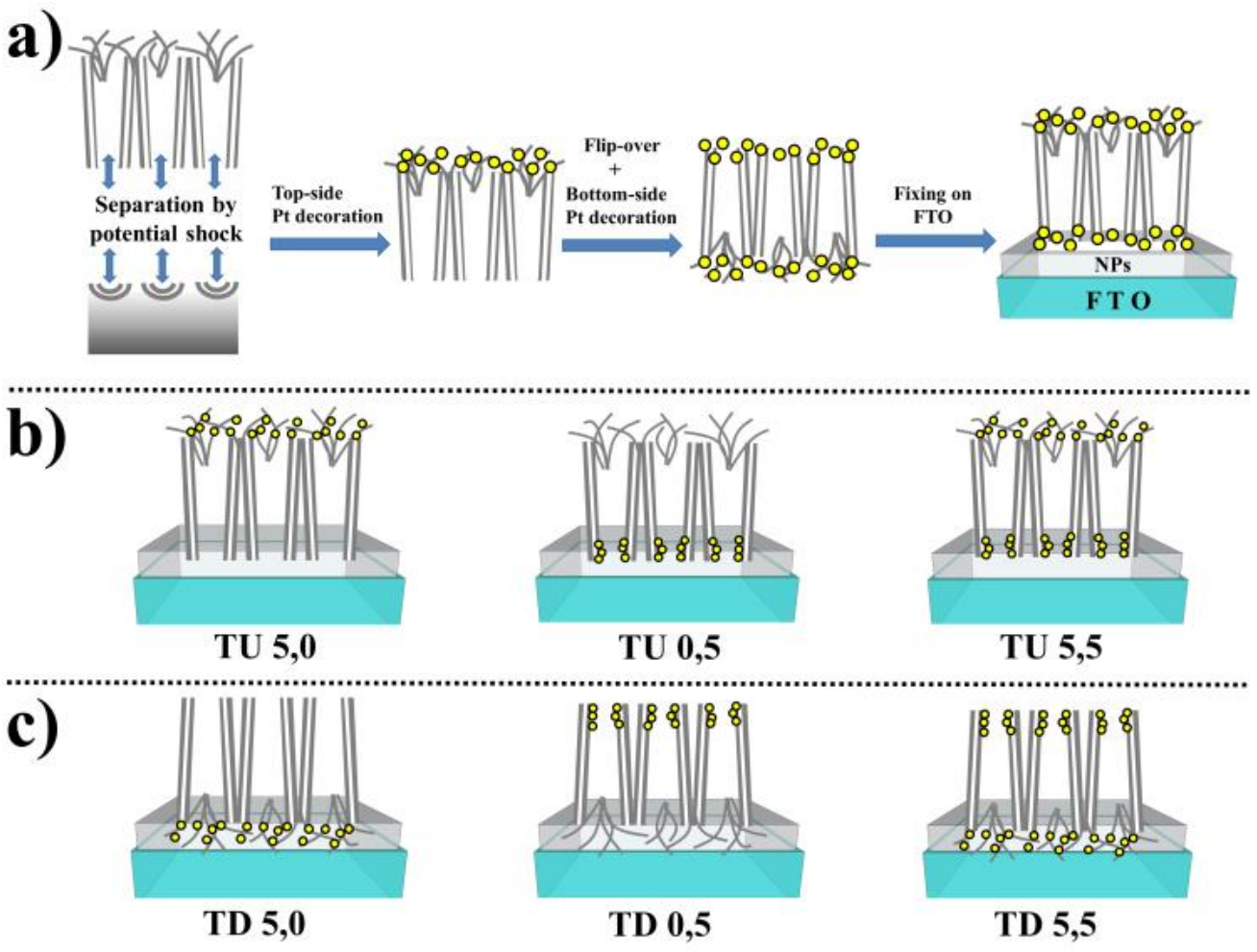

**Scheme 1.** (a) Sketch of the process used to fabricate the Pt-decorated $TiO_2$ nanotube membranes on FTO glasses; Examples of various Pt/$TiO_2$ assemblies explored as photocatalysts for $H_2$ generation in (b) tube top up and (c) tube top down configurations.

Compared to NT layers on Ti substrates, self-standing NT membranes offer some remarkable advantages. Firstly, their bottom side is accessible (and in this case open) for *e.g.* functionalization, *i.e.*, the membranes can be decorated by a single Pt sputter-coating step only at the tube top or tube bottom. Alternatively, a sequential sputtering strategy as illustrated in Scheme 1a can be used to decorate the membranes at both tube top and bottom.

Another advantage is that after cocatalyst deposition, the Pt/$TiO_2$ NT membranes can be transferred onto transparent FTO substrates. The transparent substrate allows not only for a classic direct irradiation of the photocatalytic films (*i.e.*, frontside irradiation) but also for irradiation through the FTO slide (*i.e.*, backside irradiation).

We explored various parameters, namely the nominal amount of sputtered Pt (1,3, 5 or 10 nm), the tube configuration (tube top up or tube top down), the location of the Pt decoration (only at the top, only at the bottom, at both top and bottom), and the light irradiation configuration (front or backside irradiation), and show how they affect the photocatalytic $H_2$ evolution efficiency.[51]

In order to clearly identify the different samples and provide thorough description of their structural features and configuration used in the photocatalytic runs we used the following labelling: *TU/D X,Y F/B*. *TU* and *TD* stand for “top up” and “top down”, respectively, and indicate if the tube membranes are transferred onto the FTO slides in tube top up or tube top down configuration. *X* and *Y* refer to the nominal thickness of the sputter-coated Pt film at the top and bottom of the tubes, respectively. *F* and *B* stand for “front” and “back”, respectively, and indicate the light irradiation pathway, *i.e.*, front or backside irradiation.

Scheme 1b,c illustrate examples of different photocatalyst configurations explored in this work, and Table 1 provides a description of each tube membrane configuration, amount and location of Pt cocatalyst, and light irradiation pathway.

**Table 1.** Tube configuration, nominal amount and location of Pt cocatalyst, and light irradiation pathway for various Pt/$TiO_2$ NT membranes supported by FTO slides.

| Sample name | Tube configuration<br>TU = Top up<br>TD = Top down | Pt amount / nm (nominal thickness of the sputtered film) | | Irradiation configuration<br>F = front-side<br>B = back-side |
|---|---|---|---|---|
| | | At the tube top | At the tube bottom | |
| ***TU 1,1 F*** | TU | 1 | 1 | F |
| ***TU 3,3 F*** | TU | 3 | 3 | F |
| ***TU 5,0 F*** | TU | 5 | 0 | F |
| ***TU 0,5 F*** | TU | 0 | 5 | F |
| ***TU 5,5 F*** | TU | 5 | 5 | F |
| ***TU 5,0 B*** | TU | 5 | 0 | B |
| ***TU 0,5 B*** | TU | 0 | 5 | B |
| ***TU 5,5 B*** | TU | 5 | 5 | B |
| ***TD 5,0 F*** | TD | 5 | 0 | F |
| ***TD 0,5 F*** | TD | 0 | 5 | F |
| ***TD 5,5 F*** | TD | 5 | 5 | F |
| ***TD 5,0 B*** | TD | 5 | 0 | B |
| ***TD 0,5 B*** | TD | 0 | 5 | B |
| ***TD 5,5 B*** | TD | 5 | 5 | B |
| ***TU 10,0 F*** | TU | 10 | 0 | F |
| ***TU 0,10 F*** | TU | 0 | 10 | F |
| ***TU 10,10 F*** | TU | 10 | 10 | F |
| ***TU 10,0 B*** | TU | 10 | 0 | B |
| ***TU 0,10 B*** | TU | 0 | 10 | B |
| ***TD 10,0 F*** | TD | 10 | 0 | F |
| ***TD 0,10 F*** | TD | 0 | 10 | F |
| ***TD 10,0 B*** | TD | 10 | 0 | B |
| ***TD 0,10 B*** | TD | 0 | 10 | B |

The typical structure of pristine and Pt-decorated $TiO_2$ NT membranes is shown in the SEM images in Fig. 1 (and Fig. S1).

The SEM image in Fig. 1a shows the top morphology of a pristine $TiO_2$ NT membrane. The left inset shows that a typical membrane used in this work is ~ 15 µm-thick. One can notice that the tube top morphology presents the so-called "nanograss"[52] along with remnants of the tube initiation layer. The morphology of the nanograss is highlighted in the right inset in Fig. 1a.[53] Nanograss forms by extended etching of the uppermost part of the tubes in the F-containing anodizing electrolyte. As a result, the tube walls become thin, may collapse against each other, and form aggregates of needle-like structures. These needle-like structures are a few µm long and a few tens nm wide.

In contrast, the bottom side of the membranes is characterized by a more regular and defined morphology (Fig. 1b). The potential shock technique was selected as it leads not only to successful detachment of entire tube layers (from the Ti foil) but also to opening of the tube bottom with virtually 100% success rate.[49]

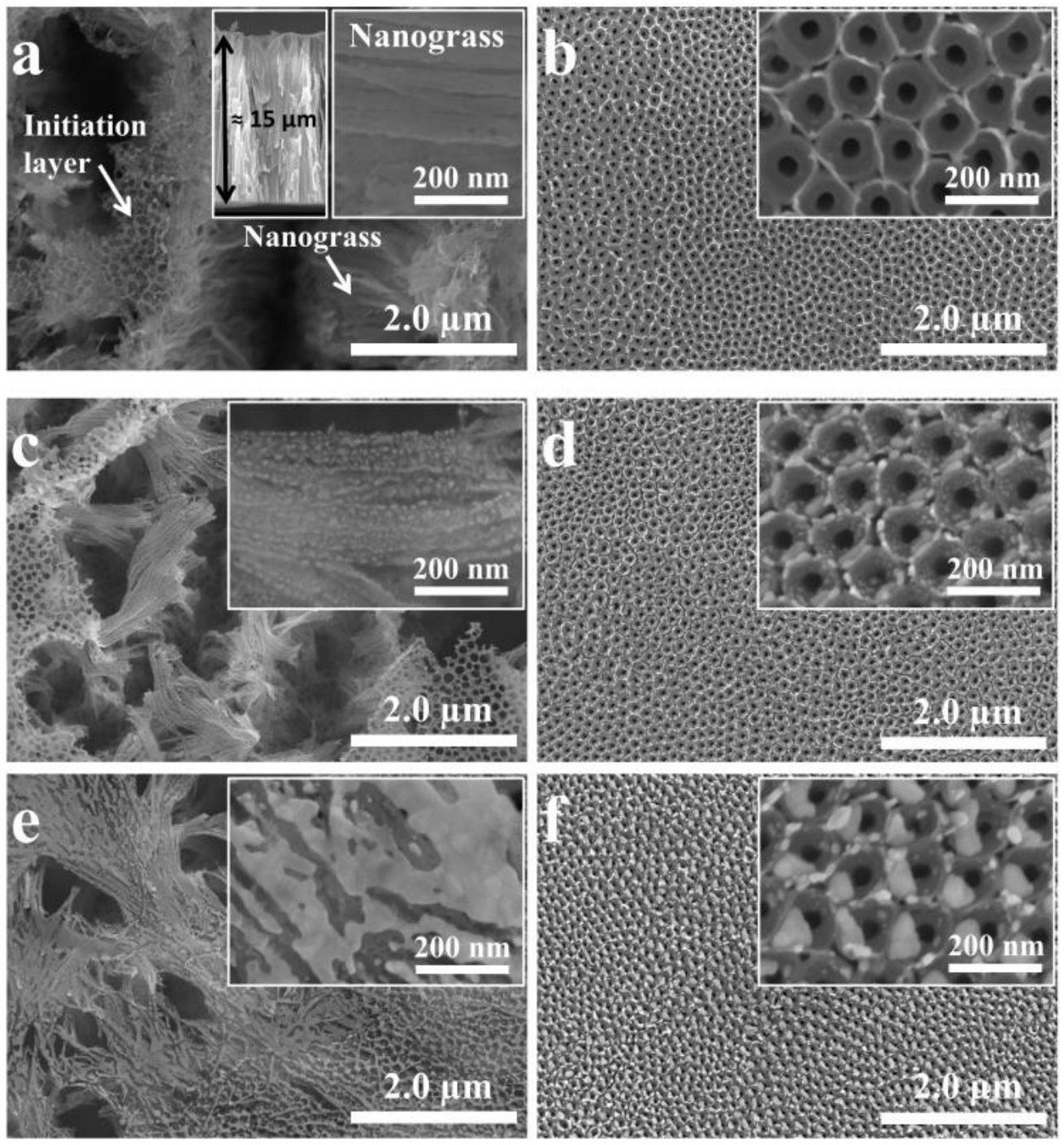

**Figure 1.** (a,c,e) Top view (insets: nanograss) and (b,d,f) bottom view SEM images of anodic $TiO_2$ nanotube membranes: (a,b) as-formed; (c,d) decorated with 5 nm-thick Pt layers at both tube top and bottom; and (e,f) decorated with 10 nm-thick Pt layers at both tube top and bottom. The left inset in (a) shows the cross-sectional SEM image of a 15 µm-thick tube membrane.

This strategy is based on applying a high potential at the end of the anodization (120 V for 3 min), which leads to local acidification and gas evolution at the oxide-metal interface and thus allows to detach the NTs with open tube bottoms.[50]

The tube bottoms are closely packed and there is almost no gap between adjacent tubes. The tube walls are ~ 60-70 nm-thick, the tube openings have an inner diameter of ~ 30-40 nm, and the tube outer diameter is ~ 100-120 nm.

The SEM images in Fig. 1(c-f) and Fig. S1 show the result of sputter-coating the NT membranes with different amount of Pt (1,3, 5 and 10 nm) at different locations (tube top or bottom).

As shown in some recent works, as-sputtered Pt films with a nominal thickness of ~ 5-10 nm (or higher) deposited on NT layers are typically conformal,[30] that is, they coat the $TiO_2$ tubular structures homogeneously. However, as-formed tube membranes are amorphous and in order to form photocatalytically active anatase $TiO_2$ membranes, the $Pt/TiO_2$ structures are annealed in air at 500°C for 1 h. Owing to the thermal treatment, and as shown in Fig. 1 (c-f) and in Fig. S1, the deposited Pt is present at the oxide surface in the form of particles and islands. This result can be ascribed to solid-state dewetting of the deposited Pt films, that is, the Pt films (when < than a critical thickness) tend to agglomerate into particles and expose the oxide surface as a result of the minimization of the free surface energy of the metal film, of the oxide substrate and of the metal-oxide interface.[54]

Particularly, the morphology of the substrate (in this case $TiO_2$ tube top and bottom) and the amount of sputtered metal largely affect size and distribution of the deposited Pt decorations,[55] as shown in Fig. S2. A first general finding that is well in line with theory on dewetting is that the thicker the metal film, the larger the dewetted particles and the broader their size distribution.[54] This means that one can control the cocatalyst particle size by adjusting the nominal thickness of the sputtered Pt film.

For 1 nm-thick Pt films (Fig. S1a,b), dewetting at tube top and bottom leads to the same results, *i.e.*, to Pt NPs with average diameter of ~ 7 nm. Also Dewetting of 3 nm-thick Pt films (Fig. S1c,d) leads to Pt NPs with average diameter of ~ 7 nm, although the size distribution becomes broader, particularly for NPs dewetted at the tube bottom.

The different morphology of the tube top compared to tube bottom leads to significant differences when sputter-dewetting Pt films > 3 nm. As shown in Fig. 1c, 5 nm-thick metal films were found to dewet at the tube top (on the $TiO_2$ nanograss) forming densely distributed spherical Pt NPs of ~ 5-10 nm in size, along with elongated Pt NPs of ~ 10-15 nm. Metal layers of the same thickness that were dewetted at the tube bottom (Fig. 1d) form spherical Pt NPs with a size distribution centered at ~ 10 nm. These NPs are found mainly at the outer perimeter of the tube bottom.

Substantially different results are obtained when 10 nm-thick Pt films are thermally treated. In this case, dewetting at the tube top (nanograss) leads to formation of voids in the Pt film. The voids have irregular shapes and are as large as several 10 up to a few 100 nm (Fig. 1e). Agglomeration of 10 nm-thick Pt films at the tube bottom leads to Pt NPs that are at ~ 20-30 and ~ 60-70 nm-sized (Fig. 1f). The bigger ones are regularly spaced (almost one Pt NP per tube opening). This results can be explained in terms of templated dewetting: the periodic geometry of the tube bottom provides a defined pathway for metal film rupture and consequent ordered agglomeration of the Pt film into NPs.[30,54,55]

Fig. 2 shows various cross sectional SEM images of Pt-decorated NT membranes. The images are taken at the very top and bottom of the tube layers. The penetration depth of the Pt NPs depends on the structure of the tube substrates.[56] The top of the tube presents an irregular geometry with relatively large voids between bundles of NTs. In this case, the sputter-deposited Pt can penetrate deep into the network of oxide nanograss – a penetration depth of up to 1 µm can be estimated (Fig. 2a,b). Conversely, the closely-packed structure of the tube bottom leads to the formation of Pt NPs mainly at the tube openings and the Pt penetration depth is less than 200 nm (Fig. 2c,d).

The XRD patterns in Fig. 3 show that upon annealing the various structures underwent crystallization into pure anatase, as demonstrated by the appearance of the main anatase reflections at 25.2° (corresponding to the (101) crystallographic plane).

The XRD data in Fig. 3a show no Pt peak for the sample *TU 0,10*. In this case Pt is selectively deposited only at the bottom of the 15 µm-thick tube layer and, owing to the configuration of the membrane on the FTO (tube top up configuration), the noble metal cannot be detected by XRD. Differently, the diffraction patterns of samples *TU 5,5* and *TU 10,0* show the typical Pt reflections at 39.7° (corresponding to the (111) crystallographic plane) – note in fact that in these samples Pt was deposited at the tube top.

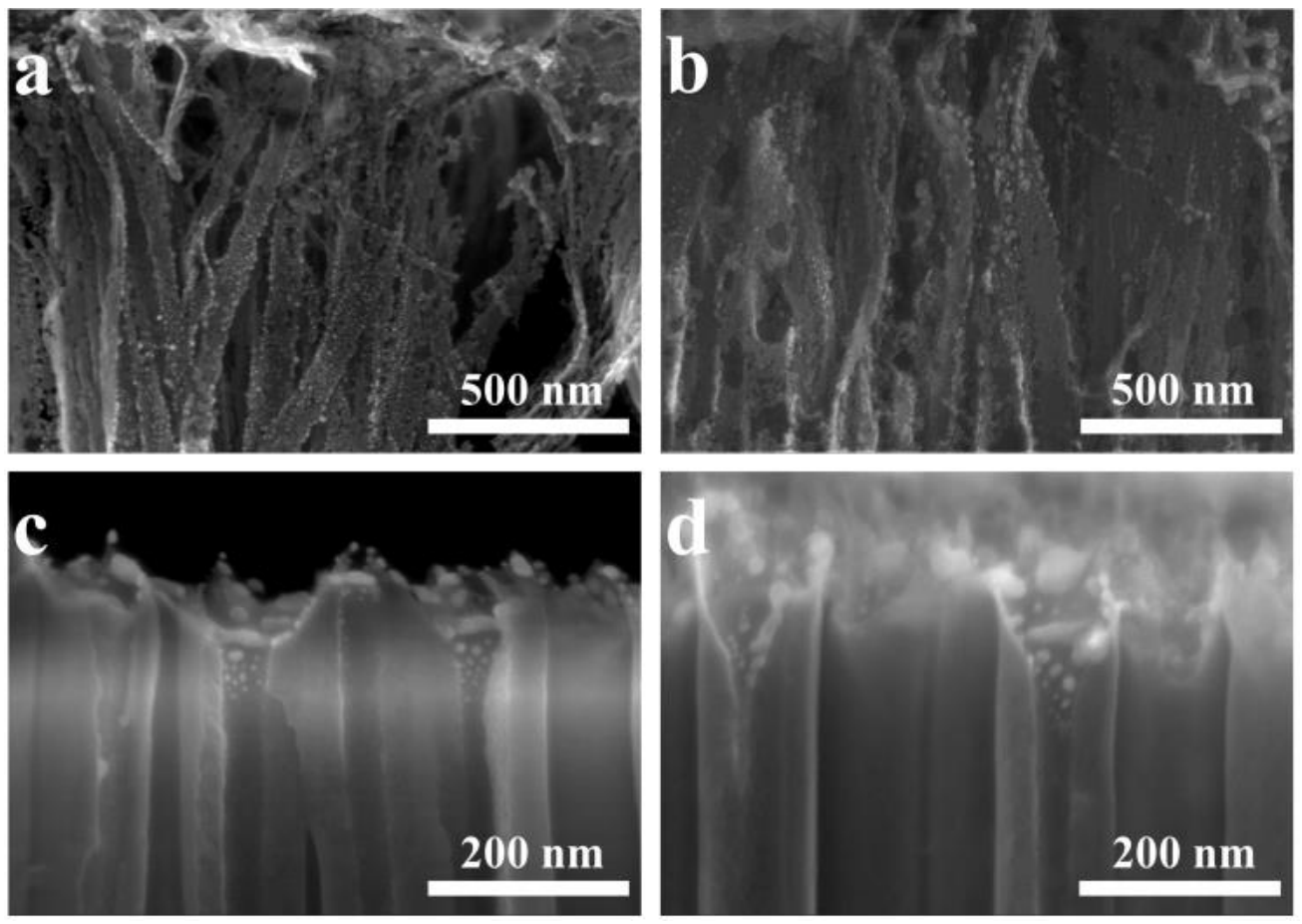

**Figure 2.** Cross sectional SEM images of NT membranes decorated with Pt NPs (a,b) at the tube top and (c,d) at the tube bottom. The Pt decoration are obtained by sputtering onto the tubes different nominal Pt amounts: (a,c) 5 nm and (b,d) 10 nm.

The intensity of the XRD Pt peak is in line with the nominal sputtered amount, that is, a clearly more intense Pt signal is observed when depositing 10 nm-thick Pt films (*TU 10,0*) compared to 5 nm (*TU 5,5*). Similar conclusions can be drawn from the XRD data in Fig. 3b for samples fabricated with membranes in tube top down configuration. This confirms that the control over amount and site-selective deposition of Pt can be achieved for any configuration.

TEM analysis (data for sample *5,5* are compiled in Fig. 3c,d) confirmed the results from both SEM and XRD investigations, that is, Pt NPs of ~ 10-15 nm are (only) found at the top (nanograss) and bottom of the $TiO_2$ NTs. At the tube bottom, bigger particles (size up to ~ 30 nm) are also visible. A HR-TEM image of Pt particles formed at the tube top (inset in Fig. 3c) shows lattice planes with distances of 3.43 and 2.00 Å, corresponding to lattice constant of anatase $TiO_2$ (101) and Pt (200) crystallographic planes, respectively.[57,58]

The Pt/$TiO_2$ membranes were further characterized by XPS, EDX, and ICP-OES in view of their chemical composition and morphology.

XPS results (Fig. 4a,b) showed the structures to be composed of Ti, O and Pt, with a minor content of N and C ascribed to remnants of the anodizing electrolyte (in the case of C there may be a contribution also from adventitious carbon).

The high-resolution *spectrum* in Fig. 4b shows for sample *TD 5,5* the typical Pt4f doublet that confirms not only the presence of the noble metal at the bottom of the tubes but also that Pt is in its metallic state $Pt^0$ (though the formation of minor amounts of Pt oxides ascribed to the thermal treatment in air at 500°C is not excluded – we have investigated this in a recent work[59]).[60]

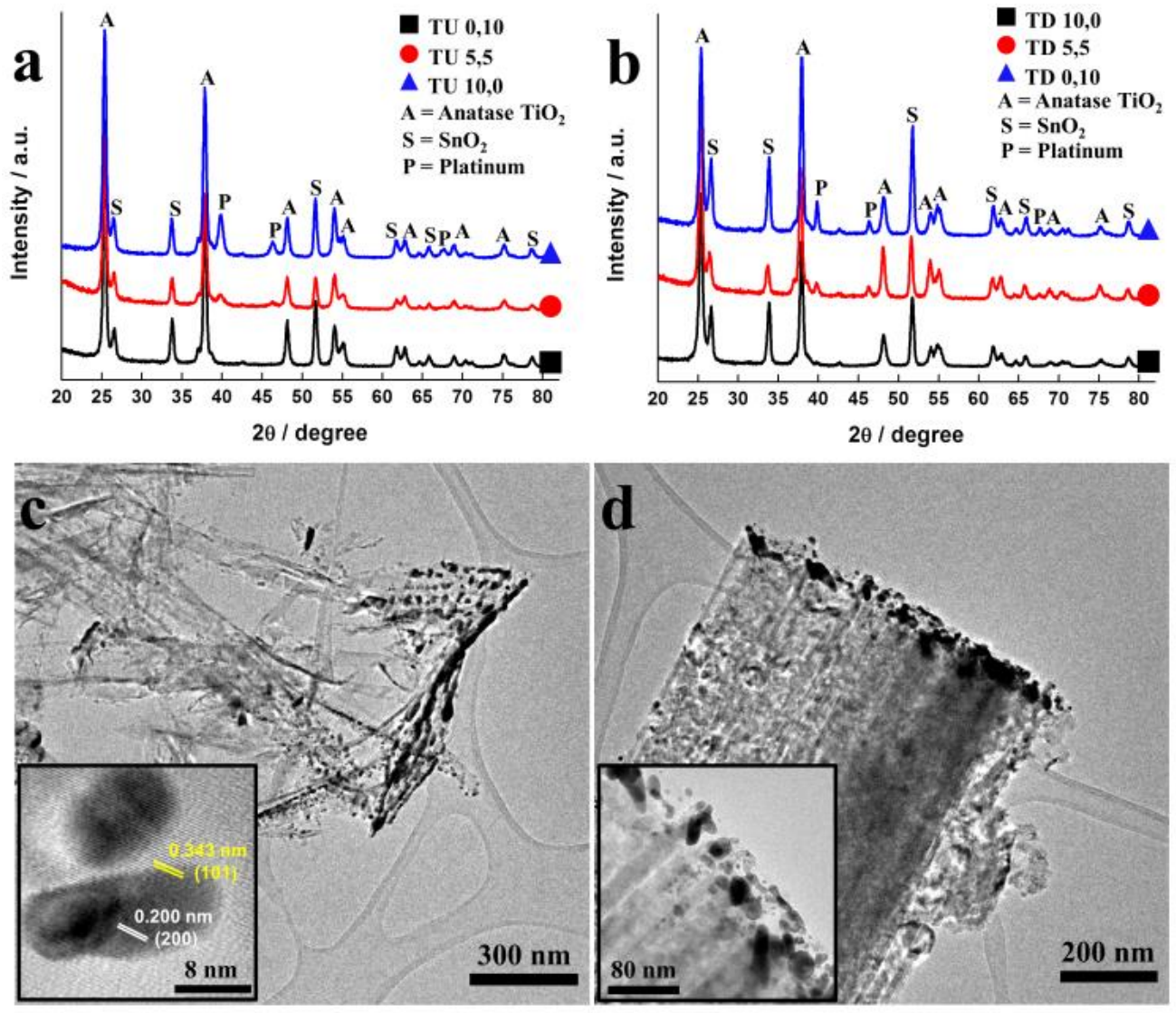

**Figure 3.** (a,b) XRD patterns of $TiO_2$ NT membranes decorated with Pt (various cocatalyst amounts and placements) and transferred onto FTO glass. The membranes are in (a) tube top up and (b) tube top down configuration. (c,d) TEM images of sample *5,5* showing the Pt-decorated (c) top and (d) bottom of the tubes. The inset in (c) shows the lattice parameters of anatase $TiO_2$ and Pt.

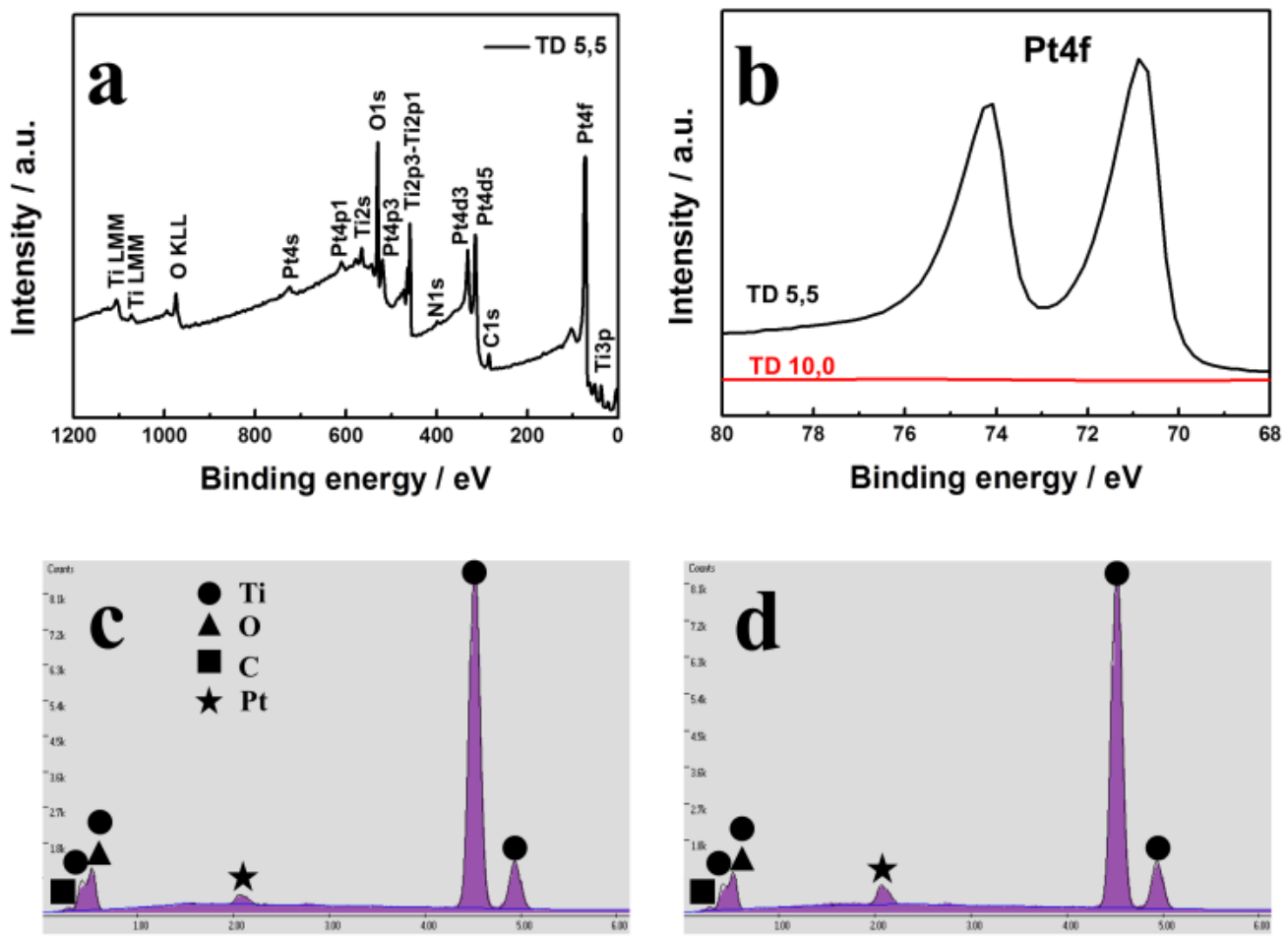

**Figure 4.** (a,b) XPS and (c,d) EDX data of various $Pt/TiO_2$ tube membranes. (a) XPS survey of sample *TD 5,5*; (b) Pt4f high-resolution spectra of samples *TD 5,5* and *TD 10,0*. EDX spectra of samples (c) *TD 5,5* and (d) *TD 0,10*.

The absence of the Pt4f doublet for the sample *TD 10,0* confirms that Pt was site-selectively deposited only at the tube top and the exposed oxide surface (*i.e.*, bottom of the tube membrane) is cocatalyst free.

The EDX *spectra* of samples *TD 5,5* and *TD 0,10* (shown in Fig. 4c,d) confirm that the structures are composed of mainly Ti and O, with small amounts of Pt and C. As for XPS data, the C content is probably ascribed to carbonaceous remnants of the anodizing electrolyte (ethylene glycol) adsorbed on the tube oxide surface[61] (or to adventitious C).

Pt content data of the different structures obtained by EDX analysis are compiled in Table 2. The measured Pt content is well in line with the nominal deposited noble metal amount. When the membranes were sputter-coated with 5 and 10 nm-thick noble metal films (at the tube bottom), the Pt content was 0.5 and 1.1 at%, respectively.

**Table 2.** Pt content evaluated from EDX analysis of various samples.

| Sample | Pt content / at% |
|---|---|
| ***TU 5,5*** | 0.5 |
| ***TU 10,5*** | 1.1 |
| ***TD 5,5*** | 0.5 |
| ***TD 0,10*** | 1.1 |

EDX-SEM cross sectional mapping was also acquired for sample *5,5*. The results (Fig. 5) show the $TiO_2$ NT layers to be composed mainly of Ti and O, with small amounts of C and F. These elements are homogeneously distributed throughout the tube layer. On the contrary, Pt is decorated only at the top and bottom extremities of the tubular structures, as shown by the intense signals in the line scan (Fig. 5d).

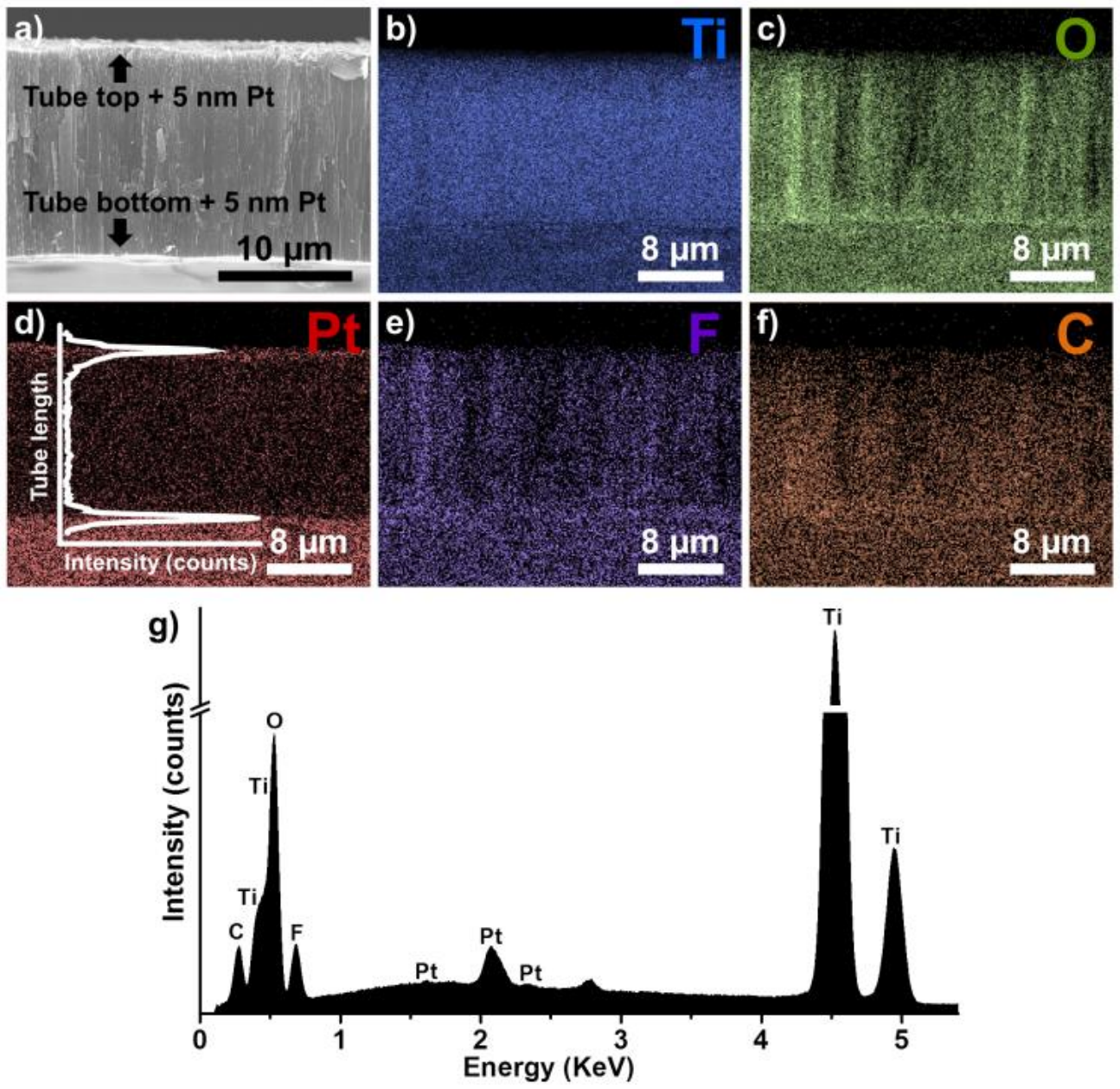

**Figure 5.** (a) SEM cross-sectional image for sample *5,5* (*i.e.*, NT membrane decorated with 5 nm Pt both at top and bottom); (b-f) EDX-SEM cross-sectional mapping for different elements: (b) Ti, (c) O, (d) Pt, (e) F and (f) C; (g) EDX spectrum measured for the cross-section of the Pt-decorated tube membrane shown in (a).

In order to determine the amount of Pt deposited on the different NT membranes ICP-OES measurements were performed for samples *1,1*, *3,3*, *5,5*, and *10,10* (see the SI for experimental details).

The results (Fig. 6) show a linear correlation between the nominal amount of sputtered Pt and the amount of Pt deposited on the $TiO_2$ NT layers. The Pt loading for the different samples (Table 3) was also determined and was in any case always ≤ ~ 0.7 wt.%, which is well in line with typical co-catalyst loadings. The highest Pt loading of ~ 0.7 wt.% is measured for a NT membrane decorated with 10 nm of Pt at both sides.

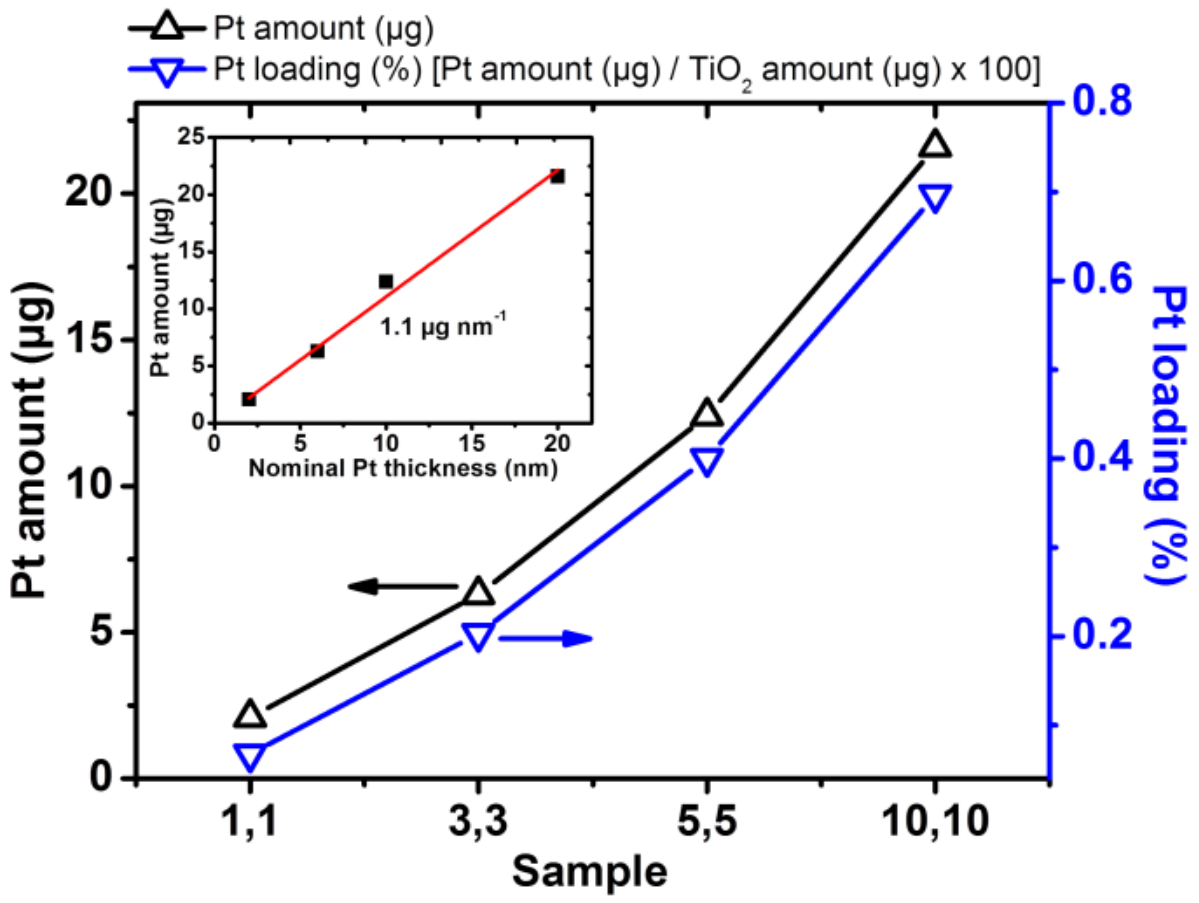

**Figure 6.** Pt amount (µg) and Pt loading (wt.%) measured by ICP-OES for different Pt-decorated tube membranes. The Pt loading is determined considering that each $TiO_2$ NT membrane has a weight of 3.1 mg. Inset: Pt amount (µg) as a function of the nominal thickness of the sputter-deposited Pt layers (*i.e.* sum of the nominal thickness of the Pt layers sputter-deposited at tube top and bottom).

**Table 3.** Pt content and loading for different Pt-decorated tube membranes determined by ICP-OES measurements.

| Sample | Nominal Pt Amount (nm)[a] | Measured Pt Amount (µg) | Pt loading (%) $\left[\frac{Pt\ (\mu g)}{TiO_2\ (\mu g)} \times 100\ (\%)\right]$ |
|---|---|---|---|
| ***1,1*** | 2 | 2.1 | 0.068 |
| ***3,3*** | 6 | 6.3 | 0.20 |
| ***5,5*** | 10 | 12.4 | 0.40 |
| ***10,10*** | 20 | 21.6 | 0.70 |

[a] The nominal Pt amount is given as the sum of the nominal thickness of the Pt layers sputter-deposited at tube top and bottom.

Overall these results confirm that Pt sputter-coating is a reliable strategy to site-selectively deposit specific amounts of noble metal cocatalyst on $TiO_2$ nanotube membranes.

The Pt/$TiO_2$ nanotube membranes were tested as photocatalyst for $H_2$ generation from water-methanol solutions under UV light irradiation (λ = 325 nm). A sacrificial hole capture agents (methanol) is added to the electrolyte to enhance the overall $H_2$ production rate.[17]

The photocatalytic results of structures fabricated with membranes in tube top up and top down configuration are compiled in Fig. 7 and Fig. 8, respectively. In general, the amount of evolved $H_2$ was found for all the samples to increase linearly over the irradiation time. Thus, the photocatalytic results can be discussed in terms of (constant) rate of $H_2$ generation (*i.e.*, $r_{H_2}$ expressed as $\mu L_{H_2}\ h^{-1}\ cm^{-2}$) obtained from the linear fitting of the $H_2$ amount *vs.* irradiation time plots.

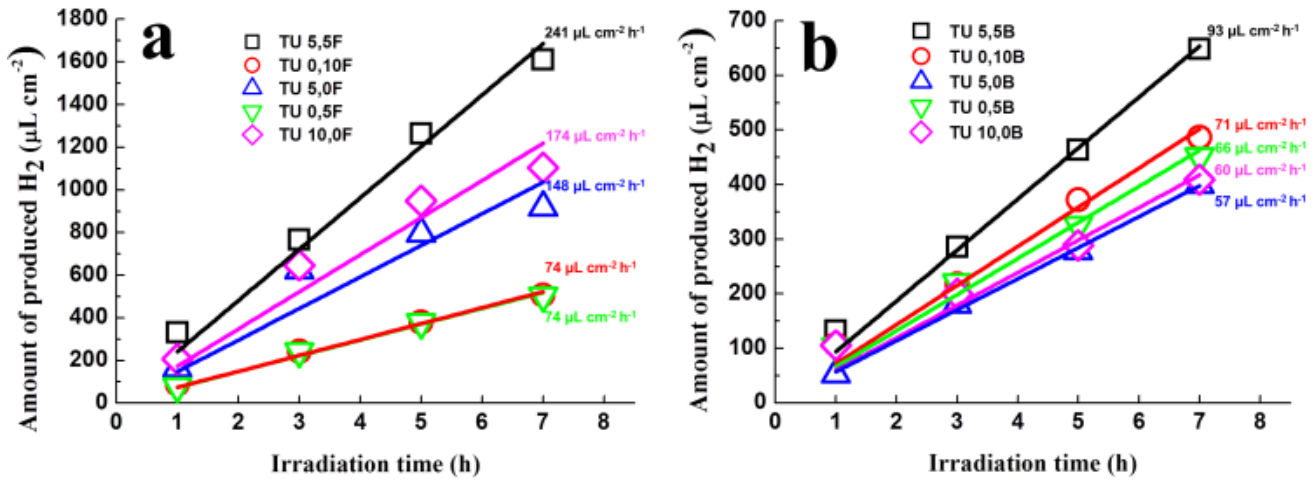

**Figure 7.** Photocatalytic results in terms of amount of generated $H_2$ *vs.* irradiation time of Pt-decorated NT membranes in tube top up configuration measured under (a) front- and (b) back-side illumination. The solid lines are linear fitting of the data for each sample.

While the $r_{H_2}$ of Pt-decorated $TiO_2$ membranes under optimized conditions can be up to *ca.* 240 µL $h^{-1}$ $cm^{-2}$, that of pristine tube membranes measured under front or back-side illumination is of ~ 3.4 and 1.5 µL $h^{-1}$ $cm^{-2}$, respectively. Thus, the presence of Pt at the $TiO_2$ surface can lead (under front-side illumination) to a 70 times higher $r_{H_2}$ than for the Pt free surface. This result is ascribed to *i)* the formation of Pt/$TiO_2$ Schottky junctions that leads to efficient electron trapping and limited charge carrier recombination (as confirmed also by the results of open circuit voltage decay (OCVD) measurements in Fig. S3); and to *ii)* the cocatalytic nature of Pt particles that mediate the electron transfer towards the environment ($H_2O$) and facilitate the $H^0$ recombination reaction.[6,38,41,44,45]

The Pt decoration of the membranes at both the extremities (top and bottom side decoration) leads in any case to a larger photocatalytic enhancement than only top and only bottom side decoration, regardless of the tube configuration and light irradiation pathway.

Particularly, Fig. 7a shows the photocatalytic results of Pt-decorated NT membranes in tube top up configuration measured under front-side illumination. Among all the investigated samples, *TU 5,5 F* delivers the highest $r_{H_2}$ (241 µL $h^{-1}$ $cm^{-2}$). Control experiments (data are compiled in Fig. S4) demonstrated that a nominal amount of Pt of 5 nm at each side of the membrane is in this case a most optimized cocatalyst loading.

It is interesting to compare the photocatalytic activity of this sample with those fabricated by depositing similar nominal Pt amounts (overall 10 nm) only at the tube top or only at tube bottom, *e.g.*, samples *TU 10,0 F* and *TU 0,10 F*. These samples lead to significantly lower $r_{H_2}$ of 174 and 74 µL $h^{-1}$ $cm^{-2}$, respectively. The different photocatalytic activity cannot be ascribed to a different surface area of the samples as dye-adsorption measurements showed virtually the same results for various $TiO_2$ NT membranes regardless of their configuration (*TU* or *TD*) and presence of Pt (data are in Table S1).

In a recent work we studied the optical properties of anodic $TiO_2$ nanotubes and found that the transmittance (*T%*) of monochromatic 325 nm light ($I_0$ = 25 mW) through a 15 µm-thick crystalline (anatase) $TiO_2$ membrane is < 1%.[47] We can thus assume that a tube membrane of similar thickness leads to a full absorption of the incident light.

Nevertheless, the tube membranes are porous and present a rough surface. Thus, while for a smooth and ideally flat titania layer the UV light absorption is assumed to take place in the first few 100 nm[62], the penetration of UV light in a porous film such as a $TiO_2$ nanotube arrays is deeper. We measured a transmittance of ~ 10% for a 1.8 µm-thick tube layer, and of ~ 1.8% for a 2.5 µm-thick tube layer. This means that the 325 nm light can reach a depth of more the 2-3 µm in the tube layers.[47] Additionally, anatase tubes were reported to provide an electron diffusion length in the range of several 10 µm.[63]

Therefore, it is plausible that the double-side Pt decoration generates two photocatalytically active zones for $H_2$ generation at the top and bottom of the tube membranes, while the central part of the NTs can be regarded as a light harvesting zone. The Schottky junctions at the tube extremities may in fact induce a local lowering of the semiconductor Fermi level ($E_F$). The charge transfer from the central part of the tube layer towards the Pt/oxide coupled zones would in this configuration be facilitated by the favorable electronic junction, that is, by the gradient of $E_F$ along the $TiO_2$ tube walls.[48]

In this context, the higher $r_{H_2}$ of single-side decorated samples *TU 0,5 F* and *TU 0,10 F* (for both ~ 74 µL $h^{-1}$ $cm^{-2}$) compared to pristine NT membranes (~ 3-4 µL $h^{-1}$ $cm^{-2}$) demonstrates that: *i)* the $H_2$ evolution can occur also at the tube bottom and hence the liquid phase can access deep zones of the porous membrane; and *ii)* in spite of the front-side illumination, electron diffusion towards the active Pt/$TiO_2$ coupled zone can take place over distances of 10-15 µm.

Fig. 7a provides also a comparison between the $r_{H_2}$ of membranes decorated with only 5 nm-thick noble metal films (*e.g.*, *TU 5,0 F*) with that of structures fabricated with overall 10 nm of Pt (*e.g.*, *TU 10,0 F* and *TU 5,5 F*). For these samples the values of $r_{H_2}$ are 148, 174 and 241 µL $h^{-1}$ $cm^{-2}$, respectively.

The data show that when the amount of Pt is increased from 5 to 10 nm at the same tube side (*TU 5,0 F* → *TU 10,0 F*) only a limited photocatalytic enhancement is observed (148 → 174 µL $h^{-1}$ $cm^{-2}$). In contrast, larger amount of cocatalyst can be more efficiently used if an overall Pt nominal amount of 10 nm is distributed at both tube sides (148 → 241 µL $h^{-1}$ $cm^{-2}$).

Another disadvantage of single-side decoration with relatively large amount of Pt (10 nm) is that the cocatalyst can cause significant shading effects, that is, the noble metal reflects the impinging light (see the results of UV/Vis reflectance measurements in Fig. S5) and a limited photon flux can reach the oxide semiconductor (this occurs when the light is irradiated through the Pt-decorated side).

Fig. 7b shows the photocatalytic results of Pt-decorated NT membranes in tube top up configuration measured under back-side illumination. While the sample fabricated with double-side Pt decoration (*TU 5,5 B*) leads also in this case to the highest photocatalytic efficiency, the $r_{H_2}$ data are lower than those measured under front-side irradiation (Fig. 7a). The reason is simply that under back-side irradiation a portion of the UV photon flux is absorbed by the FTO substrate. We measured that the *T%* (λ = 325 nm) of the FTO slide is ~ 18%.

The relatively low nominal irradiation power on the photocatalytic structures is thus the reason for the low $H_2$ generation rates measured in back-side irradiation configuration. Nevertheless, in such configuration there may be a contribution of the $TiO_2$ nanoparticle layer (underneath the tubes) to the overall photocatalytic $H_2$ generation as the UV light passing through the FTO glass directly irradiates the oxide NPs (this also implies that the transmitted light intensity that reaches the tube membranes is further reduced, *i.e.* absorbed, by the NP film). However, experiments carried out using Pt-decorated $TiO_2$ nanoparticle layers led to significantly lower $r_{H_2}$ compared to NT under optimized conditions (see Fig. S6 and Fig. S7).

Fig. 8a and Fig. 8b show the photocatalytic results of Pt-decorated NT membranes in tube top down configuration measured under front and back-side illumination, respectively. For both the data sets, the double-side Pt decoration of the tube membranes (samples *TD 5,5 F* and *TD 5,5 B*) led to the highest $H_2$ generation rate ($r_{H_2}$ = 163 and 178 µL $h^{-1}$ $cm^{-2}$, respectively). This is in line with data measured in tube top up configuration.

A general conclusion that can be drawn by comparing data in Fig. 7 with those in Fig. 8 is that the trend of photocatalytic activity can be summarized in relation to three different photocatalytic arrangements.

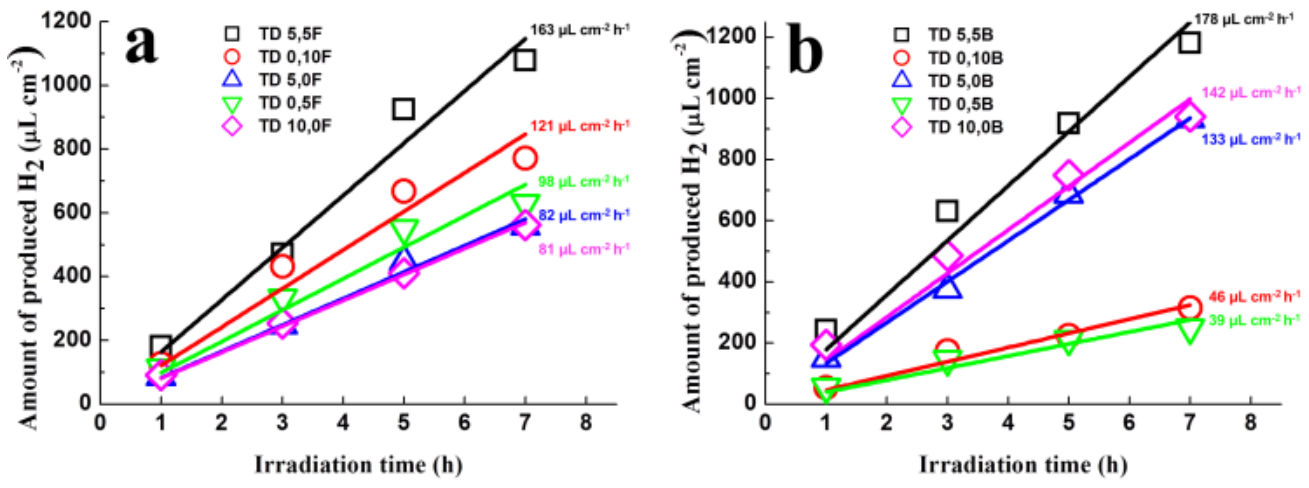

**Figure 8.** Photocatalytic results in terms of amount of generated H2 vs. irradiation time of Pt-decorated NT membranes in tube top down configuration measured under (a) front- and (b) back-side illumination. The solid lines are linear fitting of the data for each sample.

Aside from the most efficient photocatalytic configuration (*i.e.*, double-side Pt activated tubes), a slightly smaller photocatalytic enhancement is achieved when the Pt cocatalyst is deposited at the side of the membrane that is directly irradiated. In this case, the tube structure (nanograss *vs.* tube bottom) and membrane configuration (tube top up *vs.* tube top down) are also relevant factors in view of maximizing the photocatalytic activity.

The smallest increase of $H_2$ production is observed when Pt is placed at the opposite side of the membrane with respect to the illumination pathway. The poor efficiency of the latter configuration is clearly ascribed to the fact that the majority of the charge carriers generated in the $TiO_2$ NTs have to diffuse along several µm distances to reach the Pt-decorated side of the oxide membrane and react with the environment to generate $H_2$.[53] Longer diffusion pathways for the charge carriers typically lead to higher probability of electron-hole recombination and to consequent loss of photocatalytic efficiency.

In order to illustrate more in detail the benefit of a direct irradiation of the Pt/$TiO_2$ interface, some data from Fig. 7 and Fig. 8 are compiled in Fig. 9 – the $r_{H_2}$ data are of tube membranes (in various configurations) that were decorated only at one side with a 5 nm-thick Pt film and the photocatalytic configuration is such that the Pt/$TiO_2$ interface is directly irradiated.

A first remarkable finding is that regardless of the membrane configuration (top up or tube down) higher $H_2$ evolution rates were measured for membranes decorated with Pt at their top, that is, when the Pt cocatalyst was deposited at the $TiO_2$ nanograss instead of at tube bottom. This is clear when comparing the $r_{H_2}$ data of *TU 5,0 F vs. TD 0,5 F* and *TD 5,0 B vs. TU 0,5 B* (Fig. 9).

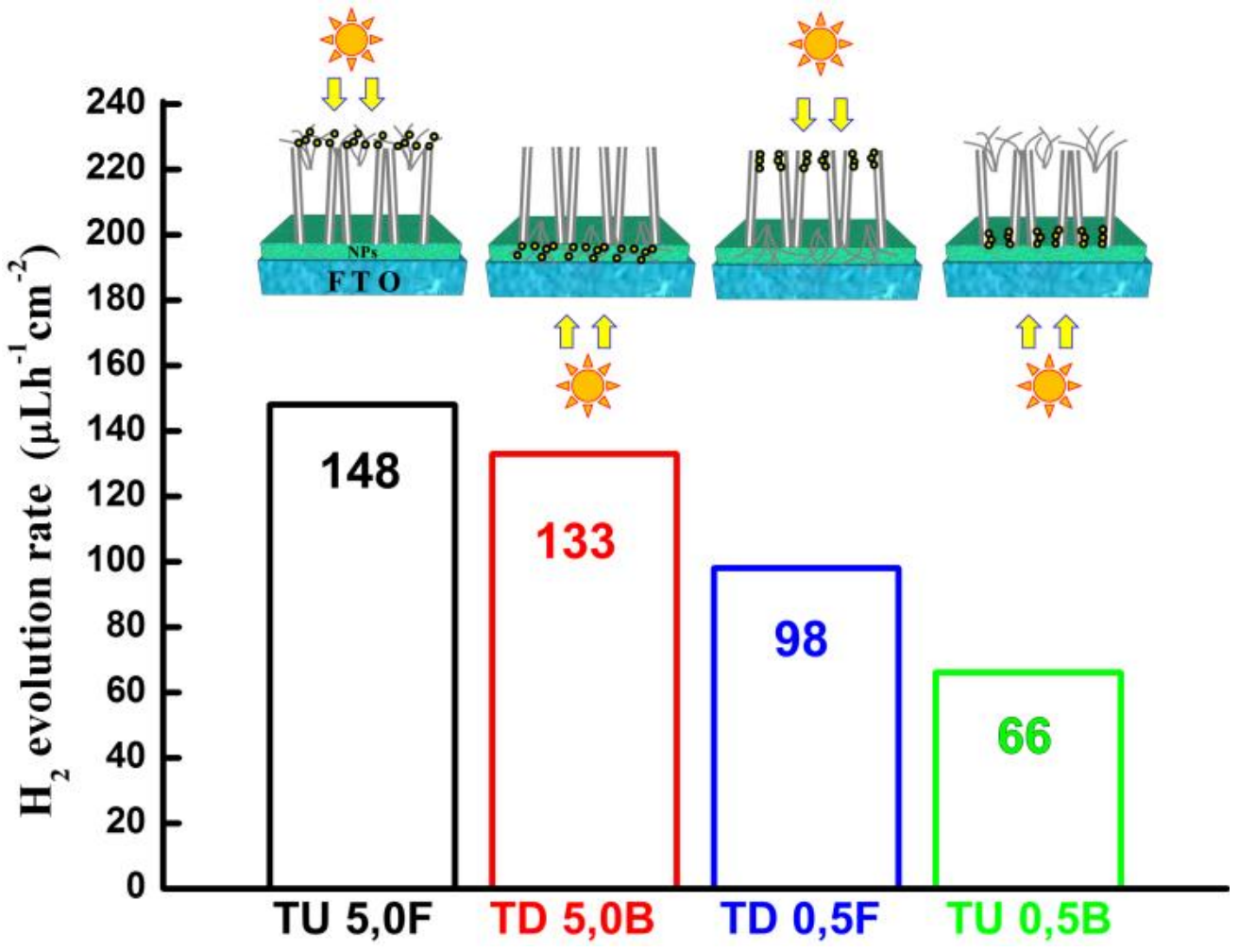

**Figure 9.** A selection of photocatalytic data from Fig. 7 and Fig. 8: the $r_{H_2}$ data are of tube membranes (in various configurations) that were decorated only at one side with a 5 nm-thick Pt film and the photocatalytic configuration is such that the Pt/$TiO_2$ interface is directly irradiated.

The reason is the different surface topography of the tube top and bottom.[56] As shown in Fig. 1, 2 and S1, the nanograss (tube top) offers a three-dimensional $TiO_2$ network with voids and empty spaces. When sputtering a 5 nm-thick Pt film over such morphology, Pt results deposited deep into the structure and homogeneously coats the uppermost 2 µm of the tube membranes. In contrast, the tube bottom offers a densely-packed morphology, where the deposited Pt penetrates only 200 nm deep into the membranes (shallow deposition) and thus the density of the Pt NPs is higher (this may induce light reflection effects and shading of $TiO_2$).

A further photocatalytic enhancement is also observed when the Pt/nanograss interface is directly exposed to the liquid phase – this is clear when comparing the $r_{H_2}$ data of *TU 5,0 F vs. TD 5,0 B* (Fig. 9).

To summarize, the photocatalytic hydrogen generation rate can be improved by using the following configuration:

- The tube membrane is decorated at both tube top and bottom with an optimized amount of Pt cocatalyst;
- The Pt-decorated membrane is transferred onto the FTO glass in tube top up configuration, therefore exposing the Pt-decorated nanograss to the environment (water-methanol solution);
- The photocatalyst is illuminated from the front-side.

## Conclusions

A critical factor needs to be optimized for efficient photocatalysis is not only the Pt amount as such, but also the cocatalyst distribution at the $TiO_2$ surface with respect to the illumination pathway. We used a photocatalytic platform for $H_2$ generation based on double-side co-catalytically activated $TiO_2$ nanotube membranes, that is, the nanotubes carry Pt particles only at the tube extremities. The photocatalytic enhancement (in terms of hydrogen generation rate) observed with this configuration can be ascribed to an optimized light absorption and electron diffusion/transfer dynamics. We propose that while the central part of the membranes act as light harvesting zone, the Pt-decorated tube extremities act as electron collection zones and mediate the charge transfer towards the reaction phase.

## Experimental Section

Ti sheets with a thickness of 0.125 mm and purity of 99.6 % (Advent Materials, UK) were used to grow the $TiO_2$ nanotube membranes. The Ti sheets were cut into 1.6 cm x 1.6 cm pieces and before anodization were cleaned in an ultrasonic bath with acetone, ethanol, and distilled water (20 min in each solvent). After sonication the Ti sheets were dried in a nitrogen stream.

The anodizing electrolyte was composed of ethylene glycol (Sigma Aldrich 99.5 %) with 0.15 M $NH_4F$ and 3 vol.% $H_2O$. The anodization experiments were conducted in a 2 electrode electrochemical cell where a Ti sheet and a Pt foil were used as working and counter electrode, respectively. 15 µm-thick $TiO_2$ nanotube layers were grown at 60 V for 1 h. The direct current (DC) bias was provided by a Jaissle potentiostat-galvanostat IMP 88PC-100V. At the end of the anodization process a potential shock at 120 V was applied for 3 min. The potential shock leads to opening of the tube bottom and double-opened self-standing tube membranes could be lifted off from the Ti substrate by soaking the samples in ethanol. A soaking time of *ca.* 20 min is generally required for complete detachment of the tube layers in the form of membranes. After detachment, the membranes were transferred on a ceramic block, coated with an FTO glass slide (to prevent curling) and dried overnight.

The NT membranes were then decorated with Pt by Ar-plasma sputtering (EM SCD500, Leica) – see Scheme 1. A Pt target (99.99%, Hauner Metallische Werkstoffe) was used as Pt source. The pressure of the sputtering chamber was reduced to $10^{-4}$ mbar and then set at $10^{-2}$ mbar of Ar during sputtering. The applied current was 16 mA. The amount of sputtered Pt, in terms of nominal thickness of the sputter-deposited Pt film, was determined *in situ* using an automated quartz crystal film-thickness sensor. The membranes were coated only at the tube top, only at the tube bottom or at both the extremities. The nominal amount of Pt deposited either at tube top or bottom was 1, 3, 5 or 10 nm.

After sputtering, the Pt-decorated $TiO_2$ NT membranes were transferred either in tube top up or in tube top down configuration onto FTO slides (TCO22-7, Solaronix) that were previously coated with a *ca*. 2.5 µm-thick $TiO_2$ nanoparticle layers (deposited by Doctor Blade method using Ti-Nanoxide HT, Solaronix). The $TiO_2$ NP layer granted mechanical adhesion of the tube membranes to the FTO slides. The Pt-decorated tube membranes supported by FTO glass were then annealed (to convert the amorphous oxide into crystalline anatase $TiO_2$) using a Rapid Thermal Annealer (Jipelec JetFirst100) at 500°C in air for 1 h, with heating/cooling rate of 30°C $min^{-1}$.

The morphology of the Pt/$TiO_2$ structures was analyzed by field-emission scanning electron microscopy (Hitachi FE-SEM S4800) and by a high resolution transmission electron microscope (TEM, Philips CM300 UltraTWIN) equipped with a LaB6 filament and operated at 300 kV.

HR-TEM analysis was performed also to determine lattice spacing of anatase $TiO_2$ and Pt. Their chemical composition was determined by energy dispersive X-ray analysis (EDAX Genesis) fitted to the Hitachi FE-SEM S4800, and by X-ray photoelectron spectroscopy (XPS, PHI 5600, US). The XPS *spectra* were shifted in relation to the C1s signal at 284.8 eV. X-ray diffraction analysis (XRD, X'pert Philips MPD with a Panalytical X'celerator detector) using graphite-monochromized Cu Kα radiation (wavelength 1.54056 Å) was used for determining the crystallographic composition of the samples.

The light transmittance of FTO was measured by irradiating the slides with monochromatic light provided by a HeCd laser (IK3552R-G, Kimmon, Japan) with emission wavelength λ = 325 nm and emission intensity $I_0$ = 60 mW cm$^{-2}$. The intensity of the monochromatic light passing through the FTO glass ($I$) was measured using a calibrated 1830-C (Newport) power meter. The transmittance was calculated as $T = \frac{I}{I_0} \times 100\%$.

For photocatalytic $H_2$ generation experiments, the Pt-decorated $TiO_2$ NT membranes on FTO slides were attached to a Ti wire. Then, these samples were immersed into a 20 vol.% methanol-water solution within a quartz tube that was used as a photocatalytic reactor (methanol was used as a hole scavenger[17]). The methanol-water solution (volume = 8 mL, kept under static conditions during the runs) and the cell headspace (volume = 6.8 mL) were purged with $N_2$ gas for 20 min prior to photocatalysis. Before sealing the quartz tube with a gastight cap, the Ti wire was stuck into the bottom side of the cap, in order to hold and immobilize the tube layers within the photocatalytic reactor. The light source used for the photocatalytic experiments was the same HeCd laser used for the transmittance measurements. The laser beam was expanded to a circle-shaped light spot of 1 cm in diameter, thus illuminating a sample surface of ~ 0.78 cm$^2$.

In order to determine the amount of $H_2$ photocatalytically generated under irradiation, the headspace of the quartz reactor was analyzed by gas chromatography (using a GCMS-QO2010SE chromatograph, Shimadzu). The GC was equipped with a thermal conductivity detector (TCD), a Restek micropacked Shin Carbon ST column (2 m × 0.53 mm), and a Zebron capillary column ZB05 MS (30 m × 0.25 mm). GC measurements were carried out at a temperature of the oven of 45°C (isothermal conditions), with the temperature of the injector set to 280°C and the TCD fixed to 260°C. The flow rate of the carrier gas, *i.e.*, argon, was 14.3 mL min$^{-1}$. All the experiments lasted 7 h, and the amount of evolved $H_2$ was measured after 1, 3, 5 and 7 h.

## Acknowledgements

The authors would like to acknowledge the ERC, the DFG, and the DFG cluster of excellence "Engineering of Advanced Materials", as well as DFG "funCOS" for financial support. Xuemei Zhou (WW4-LKO, University of Erlangen-Nuremberg, Germany) is acknowledged for helping with XPS analyses; JeongEun Yoo and Dr. Ning Liu (WW4-LKO, University of Erlangen-Nuremberg, Germany) are acknowledged for helping with light reflectance measurements. Imgon Hwang is acknowledged for helping with the open circuit voltage decay measurements.

**Keywords:** Anodization • $TiO_2$ Nanotubes • Pt Sputtering • Photocatalysis • $H_2$ generation